\documentclass[useAMS,usenatbib]{mn2e}
\usepackage{graphicx}
\usepackage{epsfig}

\newcommand{\sgrb}{\textrm{Sgr B2}}

\newcommand{\gev}{\textrm{GeV}}

\newcommand{\mhz}{\textrm{MHz}}

\newcommand{\epm}{e^{\pm}}
\newcommand{\degree}{^{\circ}}

\def\3EG{{3EG J1746-2851}}
\def\p0{{$\pi^0$}}
\def\1018{{$10^{18}$}}

\newcommand{\be}{\begin{equation}}
\newcommand{\ee}{\end{equation}}
\newcommand{\bea}{\begin{eqnarray}}
\newcommand{\eea}{\end{eqnarray}}
\newcommand{\emu}{\end{multline}}

\def\simlt{\lower.5ex\hbox{$\; \buildrel < \over \sim \;$}}
\def\simgt{\lower.5ex\hbox{$\; \buildrel > \over \sim \;$}}

\def\gcm3{{\rm\,g\,cm^{-3}}}
\def\ncm3{{\rm\,cm^{-3}}}

\def\>{$>$}
\def\<{$<$}

\title[Radio continuum and molecular line observations of Sgr B2: implications for cosmic-rays]{Interpretation of radio continuum and molecular line observations of Sgr B2: free-free and synchrotron emission, and implications for cosmic rays}
\author[R.J. Protheroe, J. Ott, R.D. Ekers, et al.]{R.J. Protheroe$^{1}$\thanks{E-mail:rprother@physics.adelaide.edu.au.}
J. Ott$^{2,3}$\thanks{currently a Jansky Fellow of the National Radio Astronomy Observatory}
R.D. Ekers$^{4}$
D.I. Jones$^{1,4}$
R.M. Crocker$^{5}$\\
$^{1}$Department of Physics, School of Chemistry \& Physics, University of Adelaide, South Australia 5000, Australia.\\
$^{2}$National Radio Astronomy Observatory, 520 Edgemont Road, Charlottesville, VA 22903, USA\\
$^{3}$California Institute of Technology, 1200 E. California Blvd., Caltech Astronomy, 105-24, Pasadena, CA, 91125, USA.\\
$^{4}$Australia Telescope National Facility, CSIRO, P.O. BOX 76 Epping, NSW 1710, Australia. \\
$^{5}$J.L. William Fellow, School of Physics, Monash University, Victoria, Australia.
}

\begin{document}

\date{Accepted . Received ; in original form .}

\pagerange{\pageref{683}--\pageref{692}} \pubyear{2008}

\maketitle

\label{firstpage}

\begin{abstract}
Recent ammonia (1,1) inversion line data on the Galactic star
forming region Sgr~B2 show that the column density is consistent
with a radial Gaussian density profile with a standard deviation
of 2.75~pc.  Deriving a formula for the virial mass of spherical
Gaussian clouds, we obtain $M_{\rm vir} =1.9 \times
10^6$~M$_\odot$ for Sgr~B2.  For this matter distribution, a
reasonable magnetic field and an impinging flux of cosmic rays of
solar neighbourhood intensity, we predict the expected
synchrotron emission from the Sgr B2 giant molecular cloud due to
secondary electrons and positrons resulting from cosmic ray
interactions, including effects of losses due to pion production
collisions during diffusive propagation into the cloud complex.

We assemble radio continuum data at frequencies between 330 MHz
and 230 GHz. From the spectral energy distribution the emission
appears to be thermal at all frequencies.  Before using these
data to constrain the predicted synchrotron flux, we first model
the spectrum as free-free emission from the known ultra compact
H{\sc ii} regions plus emission from an envelope or wind with a radial
density gradient, and obtain an excellent fit.  We thus find the
spectrum at all frequencies to be dominated by thermal emission,
and this severely constrains the possible synchrotron emission by
secondary electrons to quite low flux levels.  The absence of a
significant contribution by secondary electrons is almost
certainly due to multi-GeV energy cosmic rays being unable to
penetrate far into giant molecular clouds. This would also
explain why 100~MeV--GeV gamma-rays (from neutral pion decay or
bremsstrahlung by secondary electrons) were not observed from
Sgr~B2 by EGRET, while TeV energy gamma-rays were observed, being
produced by higher energy cosmic rays which more readily
penetrate giant molecular clouds.
\end{abstract}



\section{Introduction}

Molecular clouds have long been studied as laboratories for star
formation.  This has led to a wealth of information about the
physical characteristics of clouds and their chemical makeup,
usually obtained by observing emission/absorption lines of molecules
such as CO, OH, etc., which reveal the presence of molecular gas.
Most molecular cloud emission is thermal, from the H{\sc ii} regions and/or dust
emission.  Cosmic rays play an important role in molecular cloud
evolution by partially ionizing even the cold molecular gas,
thereby affecting, through ambipolar diffusion, the dynamics of
cloud collapse by coupling the magnetic field to the partially
ionized gas.  This in turn could result in amplification of the
ambient magnetic field during cloud collapse, and give rise to
the correlation found by \citet{Crutcher1999} between average
density of molecular gas of molecular clouds and their
line-of-sight magnetic fields.  From VLA observations of Zeeman
splitting of the  HI line \citet{Crutcher1996} found a line of sight
magnetic field for Sgr B2 of $B_{LOS}\approx 0.5$~mG to apply to
the outer envelope of the cloud complex, and
this would suggest that the magnetic field amplitude could be even higher in the inner
parts of the complex.  In fact, both \citet{Lis1989} and
\citet{Crutcher1996} actually countenance average magnetic field
strengths as high as $\sim$2 mG for the Sgr B2 cloud on the basis
of the theoretical prejudice that the cloud be magnetically
supported against gravitational collapse; we cannot exclude
that such field strengths may actually apply on large scales in
the complex.

Recent studies of the ionization rate by \citet{Tak2006} show
that the cosmic ray ionization rate of dense molecular clouds in
the Galactic Centre (GC) region may be as much as a factor
$\sim$10 higher than in molecular clouds in the solar
neighbourhood.  Other evidence for a higher cosmic ray density in
the GC region may come from the observation of 6.4~keV iron line
emission.  Assuming that low energy cosmic rays are responsible
for heating the molecular gas, \citet{Yusef-Zadeh2007} estimated
the energy densities of cosmic rays in GC molecular clouds to
range from 19 eV~cm$^{-3}$ to $6\times10^4$ eV~cm$^{-3}$, with 51
eV~cm$^{-3}$ for Sgr B2.  This is much higher than the energy
density of cosmic rays in the solar neighbourhood which is
$\sim$1 eV~cm$^{-3}$.  Since ionisation is most effective for
low-energy cosmic ray nuclei, and because cosmic ray energy
spectra typically have an inverse power-law form, it is the
cosmic rays with kinetic energies much less than 1~GeV/nucleon
that are mainly responsible.  Hence the enhanced ionization in
the GC region could be due to an overall enhancement of the
cosmic ray flux there or to an additional, steep, low-energy
component \citep{Crockeretal2007}.

High densities of low-energy hadronic cosmic ray nuclei within
regions of dense gas will result in enhanced $\sim$100~MeV
gamma-ray emission through enhanced pion production followed by
$\pi^0\to\gamma\gamma$ decay and $\pi^\pm \to \mu^\pm \to e^\pm$
decay followed by secondary electron bremsstrahlung.  The High
Energy Stereoscopic System (HESS) recently completed a survey of
the Galactic Centre \citep{Aharonian2006}, and found a broad
scale correlation between the $\gamma$-ray emission and column
density of the molecular gas, but required the GC region cosmic
ray flux at $\sim$10~TeV to be about 10 times higher than that
observed in the solar neighbourhood.  Such an enhancement in the
GC cosmic ray flux at GeV energies was not inferred by EGRET
observations of 100~MeV gamma-rays from the central region of the
Galaxy, indeed, it was explicitly noted by
\citet{Mayer-Hasselwander1998} that no localized excess
associated with the Sgr B complex was detected by EGRET excluding
the possibility of a significantly enhanced CR density in these
clouds -- {\it in the appropriate energy range}, of course.  One
possible explanation of this, and the high ionization rate
inferred by \citet{Tak2006}, that has been suggested by
\citet{Crockeretal2007} is that in the Galactic Centre region
there may be a steep component $E^{-2.7}$ with a normalization at
$\sim$10~GeV energies comparable to that locally in order to
explain the EGRET result, and an even steeper lower energy
component to explain the high ionization rate, and finally a flat
$E^{-2.2}$ component negligible at GeV energies but giving a
tenfold increase at $\sim$10~TeV to explain the HESS data.  Such
a flat component may arise as suggested by \citet{Cheng2007} if
periodic acceleration at Sgr~A* occurs when stars are tidally
disrupted at a rate of $10^{-5}$ year$^{-1}$, and diffuse at a
distance of $\sim$500~pc before $pp$ losses steepen the spectrum
on a timescale of $\sim$$10^4$ years.  However, to also explain
the broadband radio to gamma ray spectral energy distribution of
the Sgr B region, with hadronic models they need a rather strong
average magnetic field, viz., 2.2-3.7 mG.

The same interactions of cosmic ray nuclei within regions of
dense gas which may lead to enhanced gamma-ray production (at
least at TeV energies) should produce copious secondary electrons
and positrons which may in turn produce synchrotron emission in
GMC magnetic fields which are observed to be higher than
elsewhere in the interstellar medium.  This possibility was the
motivation for the present work as well as recent observations at
1.4~GHz and 2.4~GHz of the Sgr~B2 GMC
\citep{JonesSgrB2MainCmplx2008}, and of the dense cold starless
cores G333.125-0.562 and IRAS 15596-5301
\citep{Jones_StarlessCores_2008}.  The dense cores were chosen
because, unlike Sgr~B2, they are well away from the central
region of the Galaxy and would have a cosmic ray environment
expected to be similar to that of the Solar region.  They are of
much lower mass than the Sgr~B2 GMC, and their magnetic fields
are unknown.  The non-detection of these dense cold starless
cores in non-thermal emission was used to place upper limits on
the magnetic fields of both of $\sim$0.5~mG.

 Here, we investigate whether one should expect to see radio
synchrotron emission by secondary $e^\pm$ from giant molecular
cloud (GMC) complexes.  We shall compare our predictions of the
expected synchrotron emission with our recent observations
\citep{JonesSgrB2MainCmplx2008} to draw conclusions about the
cosmic ray environment around and within the Sgr~B2 GMC.  We
chose the Sgr~B2 GMC for this study because of its large mass,
its location in the central region of the Galaxy where the cosmic
ray density may be higher than that in the solar neighbourhood,
and its high magnetic field.  This was in spite of being aware of
its complicated nature, and the likely difficulty in
disentangling non-thermal from thermal emission in this source --
our search for any comparable molecular cloud with no star
formation in the central region of the Galaxy was unsuccessful.

\section{Sgr B2 cloud complex: mass and density}

The combination of magnetic fields and secondary $e^\pm$ (and
also primary $e^-$) will lead to the emission of synchrotron
radiation from molecular clouds which, because of its relatively
steep spectrum, may show up at frequencies below which the
free-free emission from H{\sc ii} regions turns down after becoming
optically thick.  The observed flux of cosmic ray $\epm$ contains
at least $\sim$15\% positrons at 10 \gev
\citep{Grimani2002,Beatty2004}.  Given that secondary electrons
and positrons would be produced {\em in situ} inside molecular
clouds by cosmic ray nuclei, they should be sites of copious
secondary $\epm$ production.  Since the production of these
secondaries is proportional to the product of the matter and
cosmic ray densities within the clouds, there should be an
appreciable flux of synchrotron radiation at low frequencies from
molecular clouds due to secondary electrons, provided cosmic rays
can penetrate the clouds.  

$\sgrb$ is one of the largest and most complex molecular
cloud/H{\sc ii} regions in the Galaxy -- see \citet{LangPalmerGoss2008} for a discussion and review of the continuum emission measurements.  It lies near the Galactic
Centre, and we assume it to be $\sim$8.5 kpc from Earth, at a
projected distance of 100 pc from the Galactic Centre.  $\sgrb$
comprises at least four components \citep{Gordon1993}.  These
are three dense cores $\sgrb$(N), (M) and (S) and a less dense
outer envelope (OE).  The dense cores are sites of massive star
formation and have H{\sc ii} regions, ultra-compact H{\sc ii} regions (UCHII)
-- \citet{Gaume1990} have found more than 60 UCHII sources in Sgr
B2(M) alone.  X-ray sources associated with H{\sc ii} regions, X-ray
sources with no radio or IR counterparts \citep{Tagaki2002},
dense cores, embedded protostars and molecular masers
\citep{Goicochea2004} are also found.  The cores are small
($\sim$0.5 pc), warm ($\approx$45--60K), light (10$^{3}$--10$^{4}$ M$_{\odot}$), dense
(10$^{6-7}$cm$^{-3}$), and correspond to $\sim$5\% of the cloud
mass.  On the other hand, the envelope is cool ($\sim$10~K),
massive ($7.6\times10^{5}$M$_{\odot}$), and less dense
(10$^{5}$cm$^{-3}$).  It is thought that at wavelengths $\lambda
>$3mm (100~GHz), free-free emission dominates, whilst at shorter
wavelengths thermal emission from dust dominates
\citep{Gordon1993}.

There have also been radio continuum, X-ray, and recently, as we
mentioned earlier, $\gamma$-ray observations of the Galactic
Centre.  The large-scale diffuse radio emission from the GC
region has been observed at 330\mhz\ extensively using the VLA
(\citet{LaRosa2005} and references therein).
It has also been observed in hard X-rays by INTEGRAL
\citep{Neronov2005}.  There have also been \emph{Chandra} and
ASCA X-ray observations of the Galactic Centre, where the authors
argue for a reflection nebula of Compton scattered X-rays from
the Galactic Centre source Sgr A* at an earlier time
\citep{Murakami2000,Tagaki2002,Murakami2001,Fryer2006}.

\begin{figure}
\includegraphics[width=8cm]{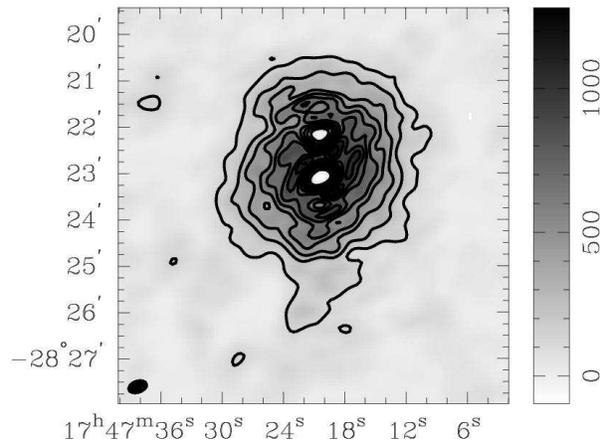}
\caption{Image and contours showing the zeroth moment of the
(1,1) line of NH$_3$ for the region around Sgr B2. J2000
coordinates are used.  Contours from 10\% peak intensity until
90\% peak intensity in increments of 10\%. Note that around the
Sgr B2(M) and Sgr B(N) H{\sc ii} regions, the NH$_3$ line
emission is strongly attenuated due to thermal bremsstrahlung
absorption. The beam is located in the lower left-hand corner of
the image, and is $26\times17''$ at a position angle of
$-70^\circ$ . The intensity scale, located on the right of the
image is from -96 to 1280 K km s$^{-1}$.  }
\label{ammonia_contours}
\end{figure}

\begin{figure}
\includegraphics[width=8cm]{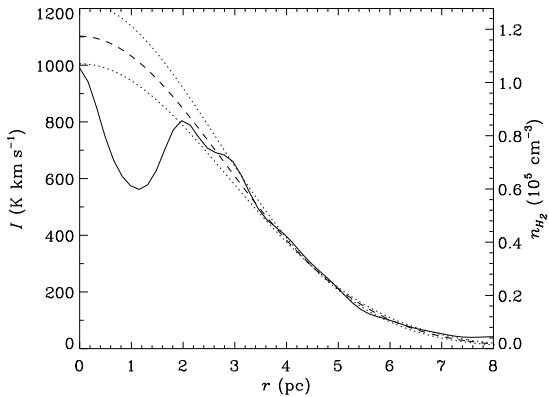}
\caption{Azimuthally averaged intensity of the (1,1) inversion
  line of NH$_3$ vs.\ impact parameter from the centre of Sgr B2
  assuming a distance of 8.5~kpc. The thick dashed line is a
  Gaussian fit to the data between 2~pc and 7~pc with
  $\sigma$=2.75~pc; at impact parameters less than 2 pc the data are affected
  by absorption effects against Sgr\,B2 (N) and (M), and beyond
  7~pc the intensity may be emission from a southern cloud
  possibly unrelated to the Sgr~B2 cloud (cf.
  Fig.\,\ref{ammonia_contours}). This fit provides the
  extrapolation of the profile toward the centre; the thin dotted
  lines give fits having $\sigma$=2.65 and 2.85~pc and are
  shown to give an indication of the uncertainty in
  $\sigma$. The right hand axis shows the inferred density as a
  function of radius for the $\sigma$=2.75~pc fit.}
\label{ammonia}
\end{figure}

Recently, \citet{Ott2006} have observed the ammonia (1,1)
inversion line over the Sgr B2 complex and the resulting zeroth
moment map (image of the intensity integrated over the line) is
shown in Fig.~\ref{ammonia_contours}. The data, obtained with the
Australia Telescope Compact Array (ATCA), show the Sgr B2 parent
molecular cloud in sharp contrast against the surrounding
molecular material. This is mainly due to the property of an
interferometer to filter out very extended structures. Excluding
the absorption of the ammonia (1,1) line by the prominent H{\sc ii}
regions Sgr~B2(M) and Sgr~B2(N), we find that the intensity
integrated over the line varies with position in such a way that
the column density profile is consistent with a two-dimensional
Gaussian with a standard deviation of $\sigma$=2.75$\pm$0.1 pc
(assuming a distance of 8.5~kpc) centred midway between Sgr~B2(N)
and Sgr~B2(M) (see Fig.~\ref{ammonia}).

For optically thin emission, the intensity integrated over the
ammonia (1,1) line is proportional to the column density of
ammonia provided the temperature is constant, and so to that of
molecular hydrogen
\[
\int_{(1,1)\rm line} I_\nu d\nu \propto N_{\rm NH_3} \propto N_{\rm H_2},
\]
where the final proportionality also assumes the fractional
NH$_3$ abundance is constant.  For the assumptions above, this
implies that the volume density profile for H$_2$ must
be a radial Gaussian density profile, too.  For the case of a cloud
with spherical symmetry, if the column density is a
two-dimensional Gaussian surface density, then the volume density
must be described by a three-dimensional Gaussian with the same
standard deviation,
\[
n_{\rm H_2}(\vec{r})={M_{\rm H_2}\over 2 m_H}{1 \over
(\sqrt{2\pi}\sigma)^3}e^{-(x^2+y^2+z^2)/(2\sigma^2)}
\]
in which case the column density at impact parameter $b$ from the cloud
centre is 
\[
 N_{\rm H_2}(b)= {M_{\rm H_2}\over 2 m_H}{1 \over 2\pi\sigma^2}e^{-b^2/(2\sigma^2)}. 
\]
Of course, the column density averages over density variations
along the line of sight, and so the smooth radial Gaussian density
profile will be an approximation to the true density
distribution.  Indeed, the cloud structure typically is fractal
and the mass distribution follows a power law, and the column
density contains contributions from a good number of individual
cloudlets. 

\subsection{Virial mass of a Gaussian spherical cloud}

Here we derive, for the first
time, the virial mass of a cloud complex with a radial Gaussian
density profile.  If a cloud is thermally supported, its kinetic energy is
\[
K =  {3 \over 2}{M \over \langle \mu \rangle m_u}kT
\]
where $\langle \mu \rangle$ is the mean atomic mass, $m_u$ is the
unified atomic mass unit, $k$ Boltzmann constant and $T$ the
temperature.  The mass inside radius $r$ of a Gaussian spherical
cloud is
\begin{eqnarray*}
M(<r) &=& M {2 \over \sqrt{\pi}} \int_0^{r^2/2\sigma^2}x^{1/2}e^{-x} dx\\ 
 &=& M {2 \over \sqrt{\pi}} \Gamma(3/2,r^2/2\sigma^2).
\end{eqnarray*}
where $\Gamma(a,x)$ is the incomplete Gamma function.  The
gravitational potential energy of a Gaussian spherical cloud is
then
\begin{eqnarray*}
U &=& - \int_0^\infty G M {2 \over \sqrt{\pi}} \Gamma(3/2,r^2/2\sigma^2){  4 \pi r^2\rho(r) \over r}dr \\
&=& -{GM^2 \over 2 \sqrt{\pi}\sigma}.
\end{eqnarray*}
Usually the temperature is obtained from observed thermal Doppler
broadening of a narrow line of some element or molecule with
atomic mass $\mu$.  Then, if the emission is optically thin, the
line has a Gaussian profile with standard deviation
(measured in m/s) of
\[
\sigma_v = \sqrt{kT \over \mu m_u}
\]
giving 
\[
 {kT \over m_u}=\mu \sigma_v^2 .
\]
From the virial theorem, $K=-{1\over 2}U$, we obtain 
\[
M_{\rm vir} = {6\sqrt{\pi}\sigma \over G}{\mu \over \langle \mu \rangle}\sigma_v^2 .
\]
Putting this in practical units, we obtain
\[
{M_{\rm vir} \over M_\odot} = 444 {\mu \over \langle \mu \rangle}\left({\sigma \over 1 \; {\rm pc}}\right)\left({v_{FWHM} \over 1 \; {\rm km\,s^{-1}}}\right)^2.
\]
This is a factor 2.1 higher than the usual formula for a uniform
density sphere of radius $R=\sigma$.

If the cloud is supported solely by turbulent motion, as is certainly the case for GMCs, the line
width is determined by the RMS turbulent velocity rather than the
thermal RMS speed of the molecular species being observed, and then
the virial mass is given by
\begin{eqnarray}
{M_{\rm vir} \over M_\odot} = 444 \left({\sigma \over 1 \; {\rm pc}}\right)\left({v_{FWHM} \over 1 \; {\rm km\,s^{-1}}}\right)^2.
\label{VirialMass}
\end{eqnarray}

\subsection{Mass of Sgr~B2 cloud complex}

The velocity FWHM of the ammonia (1,1) line observations of Sgr
B2 is 39.7 km s$^{-1}$, implying the cloud is supported by
turbulence rather than being thermally supported.  The virial
mass of Sgr B2 based on Eq.~\ref{VirialMass} and
$\sigma$=2.75$\pm$0.1 pc is $M_{\rm vir} =(1.9\pm 0.1) \times
10^6$~M$_\odot$.  Since the virial mass of a cloud complex with a
radial Gaussian density profile is a factor 2.1 higher than that
of a uniform density sphere, we suggest that the typical
uncertainty in mass determinations using the virial theorem
arising from uncertainty in cloud structure could be as large as
a factor $\sim$2.  \citet{Sato2000} gives a mass of (1--2)$\times
10^6$~M$_\odot$ for Sgr B2(M) assuming a radius of 1.5 pc, and in
the HESS paper on the GC region \citet{Aharonian2006} give a
total mass of (6--15)$\times 10^6$~M$_\odot$ for a
$0.5\degree\times0.5\degree$ = 75~pc$\times$75 pc region
surrounding Sgr B2 based on CS data \citep{Tsuboi1999}.

\begin{figure}
\includegraphics[width=8cm]{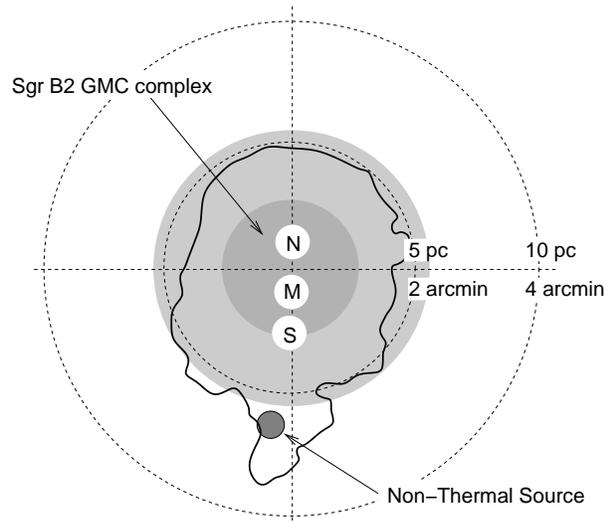}
\caption{Sketch  of the morphology of Sgr B2 showing the locations of the prominent H{\sc ii} regions, the 200~K km s$^{-1}$ NH$_3$ contour, and the locations of the main Sgr B2 cloud complex (shaded regions have radius equal to one and two standard deviations of the the assumed radial Gaussian density profile).  The strong southern non-thermal source indicated is excluded from the present analysis.}
\label{sketch}
\end{figure}

Taking the virial mass of $(1.9\pm 0.1) \times10^6$~M$_\odot$ to
be the mass of molecular gas (assumed to be almost entirely
H$_2$) in the Sgr B2 cloud complex, and a radial Gaussian density
profile with a standard deviation of $\sigma$=2.75$\pm$0.1~pc, the
maximum column density is N$_{H_2} = (2.5\pm 0.1)\times10^{24}$
cm$^{-2}$.  The density is $n_{H_2} = (1.2\pm 0.1)\times10^5$
cm$^{-3}$ at the centre of the Sgr B2 complex, and decreases to
10 cm$^{-3}$ at a radius of $\sim$12 pc which we shall consider
to be its outer radius.  The H$_2$ number density (dotted curve)
may be read off the right-hand axis in Figure \ref{ammonia}.  A
sketch of the Sgr B2 region showing the ammonia (1,1) line 10\%
contour level of the zeroth moment map, the (N), (S) and (M) H{\sc ii}
regions, and the size of inferred Gaussian cloud complex is given in
Figure \ref{sketch}.

\section{Cosmic Ray Secondary Electron Production}
The Galactic synchrotron emission is due to accelerated (primary)
cosmic ray electrons, and to electrons and positrons produced as
secondaries. The production rate of secondary electrons and
positrons depends only on the spectrum and intensity of cosmic ray
nuclei, and the density of the interstellar matter. We use the
production rate of electrons and positrons $q_m(E)$, per solar
mass of interstellar matter per unit energy (M$_\odot^{-1}$
GeV$^{-1}$ s$^{-1}$), for the cosmic ray spectrum and composition
observed above the Earth's atmosphere based on figure~4 of
\citet{Moskalenko1998}.

The production of large numbers of positrons can be traced
through their 511 keV annihilation line, and this has been
observed by INTEGRAL from within 8$\degree$ of the Galactic
Centre region \citep{Weidenspointer2006}. To explain these
observations, \citet{Beacom2006} show that the production of
positrons is $\sim10^{50}$ per year in the central region of the
Galaxy, and that they must be injected at energies below
$\sim$3 MeV to avoid excessive gamma ray emission at higher
energies (excluding the possibility that they are supplied by the
$pp \to \pi^+ \to e^+$ chain). 

The production spectrum of secondary cosmic ray e$^\pm$ has a
gradual cut-off below $\sim$0.3 GeV due to threshold for
$\pi^\pm$ in $p-p$ collisions.  Nevertheless we should check that
their production rate in molecular clouds does not exceed the
stringent constraint on production of $\sim10^{50}$ positrons per
year.  Assuming the cosmic ray spectrum in the Galactic Centre
region has the same shape there as locally, but is enhanced by a
factor $f_{CR}$, we find the total production rate of secondary
e$^+$ in Sgr B2 alone to be $1.9\times10^6$~M$_\odot f_{CR}\int
q_m(E)dE = 7.8\times10^{45} f_{CR}$ e$^+$/year.  This certainly
does not exceed the limit for $f_{CR}$ in the range 1--10 which
seems reasonable based on EGRET and HESS gamma-ray data, and
taking account of the likelihood that cosmic rays may not fully
penetrate clouds the production rate is probably less.  Indeed,
one would require $\sim3\times10^{11}M_\odot /f_{CR}$ of
interstellar gas within 1 kpc of the Galactic Centre to exceed
the observed $e^+$ production rate.

We shall delay until later in this section a discussion of the
complex problem of penetration of Galactic cosmic ray nuclei and
electrons into molecular clouds.  For the moment, then, we shall
assume that Galactic cosmic rays freely penetrate the cloud and
that their spectrum inside the cloud complex is the same as that
in the solar neighbourhood. The production spectrum of electrons
and positrons per unit volume per unit energy ($e^\pm$ cm$^{-3}$
GeV$^{-1}$s$^{-1}$) at position $\vec{r}$ in the Galactic Centre
region is then the product of the density of interstellar gas at
position $\vec{r}$ multiplied by the production rate per unit
mass
\begin{equation}
q(E, \vec{r})=f_{CR}n_{H_2}(\vec{r})q_m(E)2m_{H}/M_\odot.
\end{equation}
For moderate molecular cloud densities $n_H>10^2$ cm$^{-3}$ and
magnetic fields $B>10^{-5}$ G, the relatively short energy loss
times appear to justify neglecting diffusive transport of
electrons -- we shall discuss this point in detail in a later
section.  Then one readily obtains, by numerical integration, the
ambient number density of electrons and positrons, per unit
energy, $n^\pm(E,r)(e^\pm$ cm$^{-3}$ GeV$^{-1}$), at various
positions $\vec{r}$ within the molecular cloud complex:
\[
n(E, \vec{r}) = {\int_E^\infty q(E, \vec{r}) dE \over dE/dt},
\]
where $dE/dt$ is the total rate of energy loss of electrons at
energy $E$ due to ionization, bremsstrahlung and synchrotron
emission (because of the energies involved, we neglect positron
annihilation, and assume electrons and positrons suffer identical
energy losses).  Electrons lose energy by ionization losses in
neutral molecular hydrogen at a rate (in GeV\,s$^{-1}$) of
\[
{dE\over dt}_{\rm ioniz}=5.5 \times 10^{-17}  \left({n_{H_2}\over {\rm 1 \; cm^{-3}}}\right) \times (\ln \gamma+6.85)
\]
and by bremsstrahlung at a rate (in GeV\,s$^{-1}$) of
\[
{dE\over dt}_{\rm bremss}=1.5 \times 10^{-15}\left({E\over {\rm 1 \; GeV}}\right) \times \left({n_{H_2}\over {\rm 1 \; cm^{-3}}}\right) .
\]
The synchrotron energy loss rate  (in GeV\,s$^{-1}$) is
\[
{dE\over dt}_{\rm synch}=1.0\times 10^{-12} \times \left({B_\perp \over {\rm 1 \; gauss}}\right)^2 \times \gamma^2 
\] 
where $B_\perp$ is the component of magnetic field perpendicular
to the electron's direction. For an isotropic electron population
the solid-angle average is $\langle B_\perp\rangle= \pi B/4$, and,
assuming the magnetic field can be in any direction with respect
to the line of sight, an appropriate value for the line-of-sight
component of magnetic field obtained from Zeeman splitting
$B_{LOS}$ would be $\langle B_{LOS}\rangle= B/2$.  Hence taking
$B_\perp = \pi B_{LOS}/2$ is reasonable.  Given the observed
value is $B_{LOS}$=0.5~mG \citep{Crutcher1996}, we adopt
$B_\perp$=0.8~mG.

The synchrotron emission is calculated using standard formulae in
synchrotron radiation theory \citep{RybickiLightman1979}
\begin{eqnarray*}
j_\nu &=&  {\sqrt{3} \; e^3 \over 4\pi m_ec^2} \left({B_\perp\over {\rm \; 1 \; gauss}}\right)\times \int_{m_ec^2}^\infty F(\nu/\nu_c)n(E, \vec{r})dE\\ &&  \mbox{ ~~~ erg cm$^{-3}$ s$^{-1}$ sr$^{-1}$ Hz$^{-1}$},\\
\nu_c &=&  4.19\times 10^6 (E / m_ec^2)^2 \left({B_\perp\over{\rm \; 1 \; gauss}}\right)  \mbox{ ~~~ Hz},\\
 e &=& 4.8\times 10^{-10}  \mbox{ ~~~ esu},\\
m_ec^2 &=& 8.18\times 10^{-7} \mbox{ ~~~ erg},\\
F(x) &=& x\int_x^\infty K_{{5 \over 3}}(\xi)d\xi.
\end{eqnarray*}
and $K_{\frac{5}{3}}(x)$ is the modified Bessel function of order 5/3. 

The Razin effect reduces non-thermal emission at low
frequencies by suppression of synchrotron emission at $\nu <
\gamma_e\nu_p$ where $\nu_p$ is the plasma frequency and
$\gamma_e$ is the Lorentz factor of the radiating electrons, and
the effect is negligible where $\nu \gg 20n_e/B$ where $n_e$ is
the number density of free electrons (cm$^{-3}$) and $B$ is in
Gauss.  For the Sgr~B2 cloud assuming $B_\perp$=0.8~mG and our
lowest frequency of interest being 330~MHz, the Razin effect will
be small if $n_e \ll 10^5$~cm$^{-3}$.  Given that in a molecular
cloud $n_e \ll n_{H_2}$ and for the Sgr~B2 main complex we
estimate the central density to be $n_{H_2}=1.2\times
10^5$~cm$^{-3}$, we can safely neglect the Razin effect in the
present work.

\begin{figure}
\includegraphics[width=8cm]{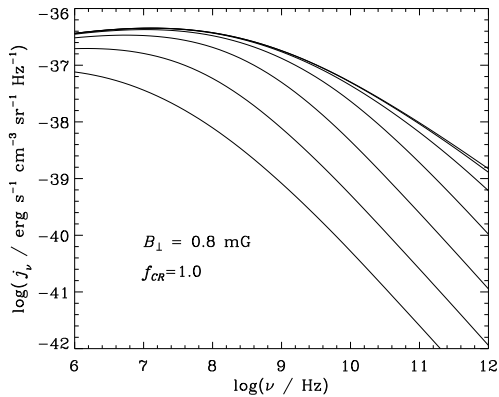}
\caption{Synchrotron emission coefficient of secondary electrons produced by cosmic ray interactions for densities $n_{H_2}=10^0$ (bottom curve), $10^1$,
$10^2$, \dots $10^6$ cm$^{-3}$ (top curve).}
\label{emission_coeff_vs_den}
\end{figure}

For a perpendicular component of magnetic field of 0.8~mG, as may
be appropriate for Sgr~B2, and the solar neighbourhood cosmic ray
spectrum, we find the specific emission coefficient due to
synchrotron emission by secondary electrons as shown in
Fig.~\ref{emission_coeff_vs_den} for various gas densities.  Note
that for higher density regions, electron energy loss by
bremsstrahlung dominates over synchrotron losses reducing the
synchrotron power per unit mass relative to lower density
regions.  As can be seen, this has the effect that at low
frequencies the synchrotron emission coefficient becomes almost
independent of density for $n_{H_2}>10^3$ cm$^{-3}$ and mG
fields.

\subsection{Penetration of Cosmic Ray Nuclei into the Sgr B2 Cloud Complex}

Work on this subject has been motivated mainly by
gamma-ray observations, particularly of the central region of the
Galaxy.  An important contribution to the Galactic gamma-ray
intensity comes from interactions of cosmic ray nuclei through
pion production and subsequent decay $\pi^0\to\gamma\gamma$, and
$\pi^\pm\to\mu^\pm\to e^\pm$ followed by bremsstrahlung or inverse
Compton. Of course primary accelerated electrons are also important
for the latter two processes.  Put simply, if cosmic rays can freely
enter molecular clouds then the gamma-ray flux will be higher than
if they cannot.  In the present work we are interested in
synchrotron radiation by the same secondary $e^\pm$.  Another
motivation has been to estimate the ionization rate due to cosmic
rays.  Again, this depends crucially on the extent of penetration
of cosmic rays responsible for ionization, mainly those of lower
energy.

The nature and extent of penetration of cosmic rays into
molecular clouds has is not yet fully understood, and has a long
history.  \citet{Skilling1976} concluded that the very low energy
cosmic rays mainly responsible for ionization of cloud material
are efficiently excluded from clouds, whereas
\citet{Cesarsky1978} concluded that molecular clouds are pervaded
by inter-cloud cosmic rays.  \citet{Dogel1990} considered
acceleration of charged particles by turbulence in giant
molecular clouds, and suggested that this mechanism may explain
the unidentified gamma-ray sources discovered by COS-B, and
estimated the synchrotron radio emission of accelerated primary
and secondary electrons in molecular clouds in this scenario.

Certainly, at multi-TeV energies there should be no problem in
cloud penetration, and this was recently confirmed by the
excellent correlation between TeV gamma-ray intensity as measured
by HESS \citep{Aharonian2006} and the column density of molecular
gas for the Galactic Centre region.  At lower energies, the
interpretation of the 100~MeV--GeV energy gamma ray intensity
measured by EGRET \citep{Mayer-Hasselwander1998} toward the
Galactic Centre region is ambiguous and we await with great
interest the higher resolution data from GLAST due for launch in
2008.  In this context, \citet{Gabici2007} have recently
investigated the penetration of cosmic rays into molecular clouds
to understand the importance of this for gamma ray emission at
GeV and TeV energies.  They considered proton-proton collision
losses as cosmic rays diffuse into a cloud.  Taking a typical
cloud to have $n_{H_2}$=300~cm$^{-3}$, $B$=10~$\mu$G and radius
20~pc, they found that, for a diffusion coefficient based on that
which seems to apply to cosmic rays throughout the Galaxy (as
determined by secondary to primary composition measurements),
cosmic rays would freely penetrate.  They also found, however,
that if the diffusion coefficient inside the cloud is smaller,
say 0.01 of the average Galactic one, that exclusion becomes
relevant for 10--100~GeV cosmic ray nuclei resulting in
suppression of GeV gamma-ray emission. Given that the Sgr~B2
complex has a much higher density and magnetic field than that
modelled by \citet{Gabici2007}, it is clearly necessary to
determine the extent of suppression of $e^\pm$ production.

Cosmic ray protons and nuclei produced outside the cloud will
penetrate the cloud by diffusion and advection, and lose typically
half their energy in \emph{pp} collisions on a time scale
t$_{pp}\approx5\times10^7 n_{H_2}^{-1}$ yr, where $n_{H_2}$ is in
cm$^{-3}$.  The advection timescale is $t_{\rm adv}\equiv R_{\rm cloud}
/\sigma_v$ which for $R_{\rm cloud}$=12 pc and $\sigma_v$=39.7 km s$^{-1}$
is $t_{\rm adv}\simeq3\times10^5$ years.  

The diffusion timescale is
\begin{equation}
	t_{\rm diff}(E)\equiv\frac{R_{\rm cloud}^2}{2D(E)}
\end{equation}
where $D(E)$ is the diffusion coefficient, which depends on the
ambient magnetic field and the spectrum of turbulence. The
minimum diffusion coefficient for a completely tangled magnetic
field is the so-called ``Bohm diffusion coefficient'' which, for
relativistic protons, is $D_{\rm
min}(E)=\frac{1}{3}r_g(E)c\propto E$ where
$r_g\approx10^{-9}E_{\rm GeV}B^{-1}_{\rm mG}$ pc is the
gyroradius, $B_{\rm mG}$ is the magnetic field in milligauss and
$E_{\rm GeV}$ is the proton energy in GeV. Models of cosmic ray
propagation in the Galaxy which are consistent with the observed
relative abundance of ``primary'' cosmic ray nuclei (e.g.\ Carbon) and
``secondary'' cosmic ray nuclei (e.g.\ Boron) -- the latter produced by
spallation of primary cosmic ray nuclei -- suggest that
$D(E)\propto E^\alpha$ where $\alpha\sim0.3-0.7$. If a Kolmogorov
spectrum of turbulence is present, then one would expect
$\alpha$= 1/3, but in the presence of strong magnetic fields a
Kraichnan spectrum may give rise to $\alpha$= 1/2. Following
\citet{Gabici2007}, we adopt a diffusion coefficient
\begin{equation}
	D(E)=3\times10^{27}\chi\left[\frac{E/(1 \
	\textrm{GeV})}{B/(3 \ \mu G)}\right]^{0.5} \
	\textrm{cm}^2 \ \textrm{s}^{-1}
\label{Eq:diffco_Gabici}
\end{equation}
where $\chi\leq$1 is a factor to account for the possible
suppression (slowing) of diffusive transport.

Adopting a magnetic field of $\sim$0.8 mG, the typical energies
of electrons or positrons responsible for synchrotron emission at
330 MHz - 1 GHz are $\sim$0.3--0.4 GeV. Taking the primary proton
energy to be $\sim$10 times higher, and setting the outer
boundary of the Sgr B2 cloud to be $\sim$12 pc, this gives a
diffusion timescale of $\sim 10^4\chi^{-1}$ yr.  Comparing this
diffusion time-scale with the loss timescale of only 500 years,
for $pp$ collisions in a central density of $n_{H_2}\sim10^5$
cm$^{-3}$, shows that, for the 0.8 mG magnetic field, penetration
to the centre of the GMC complex is very improbable below $\sim$3
GeV energies. As we have already noted, because bremsstrahlung
losses dominate in dense regions much of the synchrotron emission
is expected to come from outer regions of Sgr B2, and it is
penetration to these outer regions that matters most.  For these
regions the distance is obviously smaller and the density lower,
suggesting that penetration of the outer region of the GMC
complex is less of a problem.

We can make a more quantitative approximation of cosmic ray
nucleus penetration by analogy with scattering (`s') and
absorption (`a') of radiation and we define an effective optical
thickness $\tau_{\star}=\tau_{a}(\tau_a+\tau_s)$ analogous to
that used when considering radiative diffusion -- see, e.g.,
\citet{RybickiLightman1979}. In our case, for penetration from an
outer boundary $R$ to distance $r$ from the centre we have
\begin{equation}
	\tau_a(r)\approx\int^R_r0.5[2n_{H_2}(r')]\sigma_{pp}dr'
\end{equation}
where $\sigma_{pp}(E)\approx$35 mb above threshold, and the 0.5
factor is approximately the mean inelasticity (fractional energy lost) in $pp$
collisions, and
\begin{equation}
	\tau_s(E,r)\approx\int^R_r\frac{c}{3D(E,r')}dr'
\end{equation}
since for isotropic diffusion, the effective mean free path is
$3D/c$.  Then the cosmic ray intensity at radius $r$ is
$I_{CR}\approx e^{-\tau_\star(E,r)}I_{CR}(E,R)$.  

The mean primary proton energy for a given secondary electron energy is given by
\[
\langle E_p \rangle = {\int E_p n_p(E_p) Y(E_{e};E_p) d E_p \over \int n_p(E_p) Y(E_{e};E_p) d E_p }
\]
where $n_p(E_p)dE_p$ (cm$^{-3}$ GeV$^{-1}$) is the number density
and of cosmic ray protons with energy $E_p$ to $(E_p+dE_p)$, and
$Y(E_{e};E_p)dE_{e^\pm}$ (g$^{-1}$ cm$^{2}$) are particle yields
giving the rate of production of secondary electrons and
positrons with energy $E_e$ to $(E_e+dE_e)$ per unit pathlength
(in g cm$^{-2}$) by a single cosmic proton of energy $E_p$.
These particle yields can be obtained from accelerator data on
charged pion production in $pp$ collisions taking account of
$\pi\to\mu\to e$ decays (we use data kindly provided by
T.~Stanev, private communication).  We find that for the local
cosmic ray spectrum the mean primary proton total energy is
\begin{equation}
\langle E_p \rangle \approx 0.015 
\gamma_{e} + 22 \gamma_{e}^{-0.5} \mbox{~~ GeV}.
\label{MeanProtonEnergy_of_SecElectron}
\end{equation}
So for observations made at frequency $\nu$, the appropriate
cosmic ray proton energy to use is determined by assuming
electrons radiate at the critical frequency, i.e.\ putting
\begin{equation}
\gamma_e = \left({\nu\over4.19\times 10^6 B_\perp}\right)^{1/2} 
\end{equation}
into Eq.~\ref{MeanProtonEnergy_of_SecElectron}, 
where $\nu$ is in Hz, $B_\perp$ in gauss.  

\begin{figure}
\includegraphics[width=8cm]{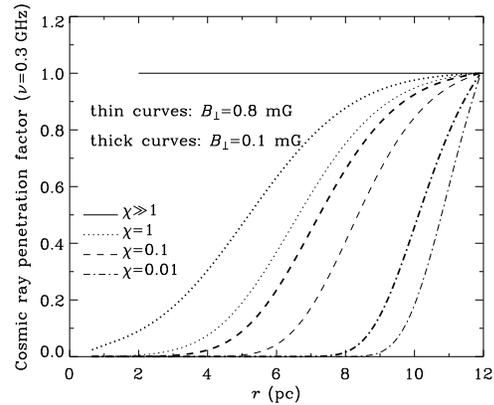}
\caption{The cosmic ray penetration factor
$e^{-\tau_*[E_p(\nu,r),r]}$ appropriate to $\nu$=0.3~GHz is plotted
against radius for the diffusion coefficient defined by
Eq~\protect\ref{Eq:diffco_Gabici} with magnetic field and $\chi$
as labeled ($\chi\gg 1$ corresponds to unimpeded penetration).}
\label{CR_penetration}
\end{figure}

The penetration factor $e^{-\tau_\star(E,r)}$ appropriate to an
observing frequency of 0.3~GHz is plotted against radius from the
centre of the Sgr~B2 GMC for various values of the diffusive
transport suppression factor $\chi$ in
Fig.~\ref{CR_penetration} for $B_\perp$=0.8~mG.  To show how
our results depend on assumed magnetic field, here and elsewhere
in this paper we shall also show results for a lower magnetic
field, arbitrarily chosen to be $B_\perp$=0.1~mG.  The
penetration factor for this magnetic field is also shown in
Fig.~\ref{CR_penetration}.  

\subsection{Diffusion of Secondary and Primary Cosmic Ray Electrons in the Sgr B2 Cloud Complex}

\begin{figure}
\includegraphics[width=8cm]{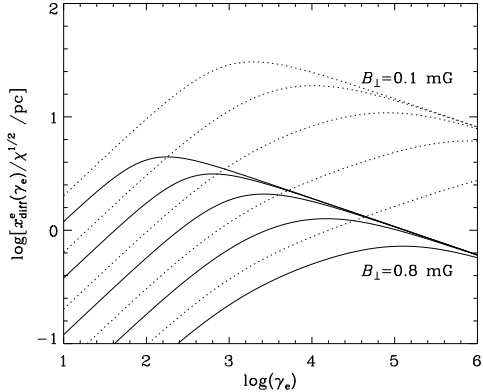}
\caption{The diffusion-loss distance as defined by
Eq.~\protect\ref{Eq:ElectronPenetration} plotted vs.\ $\gamma_e$ for the
diffusion coefficient given by Eq~\protect\ref{Eq:diffco_Gabici} for two
magnetic fields (as labelled) and $n_{H_2}$=$10^5$~cm$^{-3}$
(bottom curves), $10^4$~cm$^{-3}$ \dots $10^1$~cm$^{-3}$ (top curves).}
\label{ElectronPenetration}
\end{figure}

Before predicting the synchrotron emission from the Srg~B2 GMC,
we consider the diffusion of secondary $e^\pm$ and primary cosmic ray $e^-$.
Due to the high magnetic field, and high densities within the
inner part of the GMC, electrons will suffer rapid energy
losses there and this will limit how far they can propagate by
diffusion.  An estimate of how far an electron with Lorentz factor
$\gamma_e$ can propagate by diffusion before losing a significant
fraction of its energy is given by what we shall refer to as the ``diffusion-loss
distance''
\begin{equation}
x_{\rm diff}^e(\gamma_e) = \left[{D(\gamma_e m_ec^2)\gamma_e \over (d\gamma_e/dt)_{\rm total}}\right]^{1/2}.
\label{Eq:ElectronPenetration}
\end{equation}
We plot $x_{\rm diff}^e(\gamma_e)$ for $B_\perp$=0.8~mG and
$\chi$=1 in Fig.~\ref{ElectronPenetration}, and also for
$B_\perp$=0.1~mG to show the effect of a significantly lower
magnetic field than appears to be present over the Sgr~B2 GMC.
For lower $\chi$ values, the diffusion-loss distance is lower and
is obtained by multiplying by $\sqrt{\chi}$.

For synchrotron radiation at the adopted magnetic field,
$B_\perp$=0.8~mG, the Lorentz factor of electrons mainly
responsible for emission at 1~GHz is $\gamma_e$$\approx$550.
Near the centre of the GMC where $n_{H_2}$$\sim$$10^5$~cm$^{-3}$
the diffusion loss distance is $x_{\rm diff}^e(550)\sim$0.2~pc,
at $r$=6~pc where $n_{H_2}$$\sim$$10^4$~cm$^{-3}$ we find $x_{\rm
diff}^e(550)\sim$0.5~pc, at $r$=8~pc where
$n_{H_2}$$\sim$2$\times$$10^3$~cm$^{-3}$ we find $\sim$1.3~pc,
and at $r$=12~pc where $n_{H_2}$$\sim$$10$~cm$^{-3}$ we find
$x_{\rm diff}^e(550)\sim$3.7~pc.  From this we can draw the
following conclusions: (a) at all distances the diffusion loss
distance of electrons producing synchrotron radiation at 1~GHz is
small compared to the radial coordinate and so we may safely make
the approximation that secondary electrons within the GMC radiate
where they are produced; (b) primary cosmic ray electrons from
outside the GMC will not be able to propagate significantly
towards the centre of the GMC and so are effectively excluded
from the GMC complex.  These conclusions are made even stronger
if $\chi$$<$1.  However, primary electrons accelerated inside the
GMC, e.g.\ by diffusive shock acceleration at supernova shocks or
wind shocks, will produce synchrotron emission inside the GMC.

For the case of a weaker magnetic field, e.g.\
$B_\perp$=0.1~mG, which is lower than appears to be present in the Sgr~B2
GMC but which may occur in some other clouds, the Lorentz factor
of electrons mainly responsible for emission at 1~GHz is
$\gamma_e$$\approx$1500, and more significant penetration of
primary cosmic ray electrons would take place unless $\chi \ll
1$.

\subsection{Predicted synchrotron emission from Sgr B2}

\begin{figure}
\includegraphics[width=8cm]{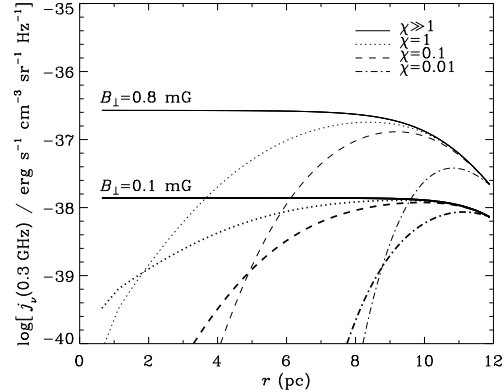}
\caption{Emission coefficient at frequency $\nu$=0.3~GHz vs.\
distance from the centre of the Sgr~B2 GMC for $B_\perp$ as indicated and for diffusion as  defined by
Eq~\ref{Eq:diffco_Gabici} with $\chi$ as labeled.}
\label{emission_coeff_vs_r}
\end{figure}

At frequency $\nu$, for each point within the Sgr~B2 giant
molecular cloud complex we determine the H$_2$ number density to
find the synchrotron emission coefficient corresponding to
complete penetration of cosmic rays within the cloud complex.
Multiplying this by the cosmic ray penetration factor and
$f_{CR}$ we obtain the synchrotron emission coefficient
$j_\nu(\vec{r})$ taking account of cloud penetration and the
possibility of cosmic ray enhancement in the Galactic Centre
region compared to that locally.  The synchrotron emission
coefficient at 0.3~GHz is plotted as a function of distance from the centre
of the Sgr~B2 GMC for $B_\perp$=0.8~mG and
$B_\perp$=0.1~mG in Fig.~\ref{emission_coeff_vs_r}.  In both
cases, for $\chi \le 1$ the emission coefficient at the cloud
complex centre is negligible compared to that near its edge.

\begin{figure}
\includegraphics[width=8cm]{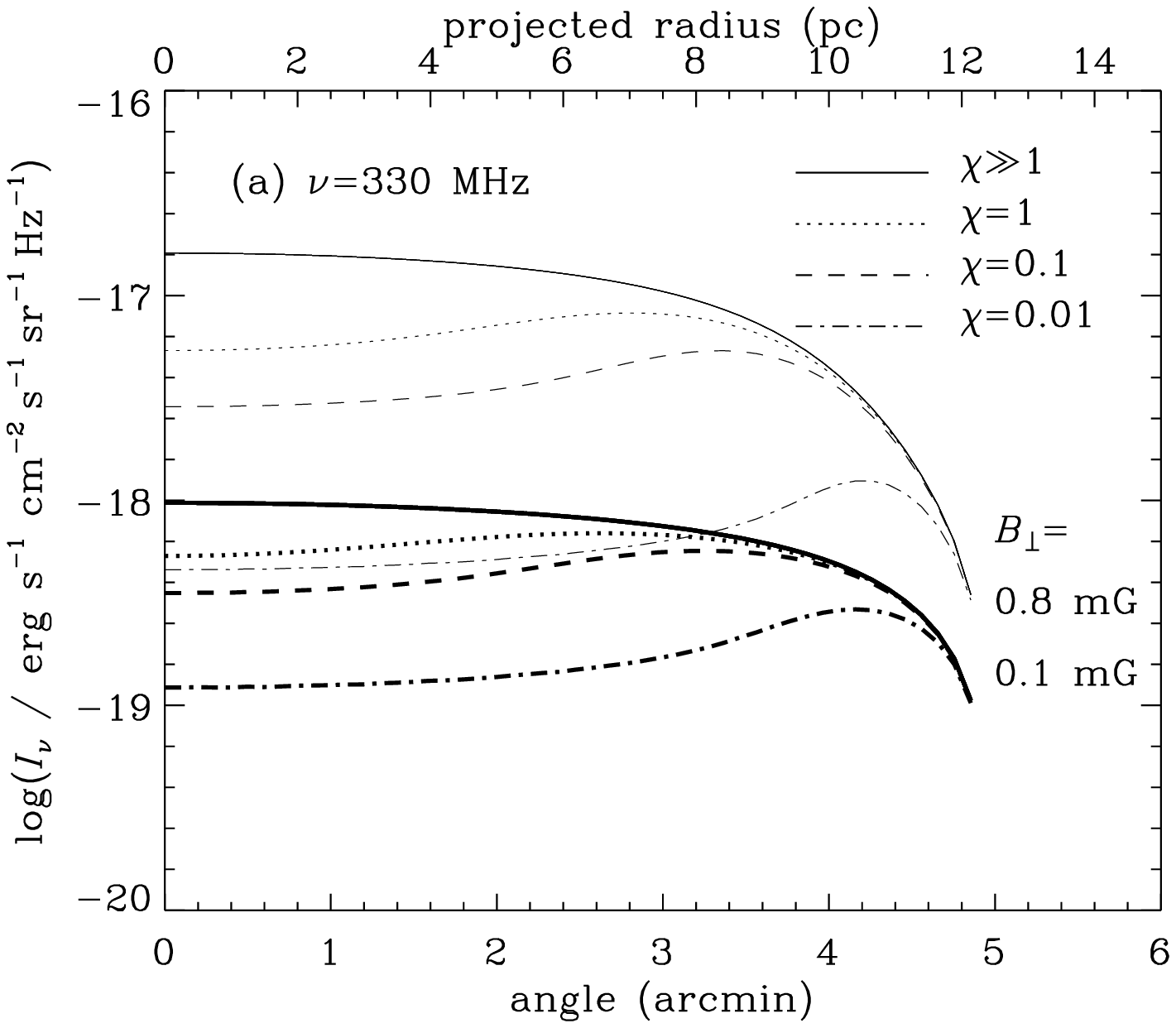}

\includegraphics[width=8cm]{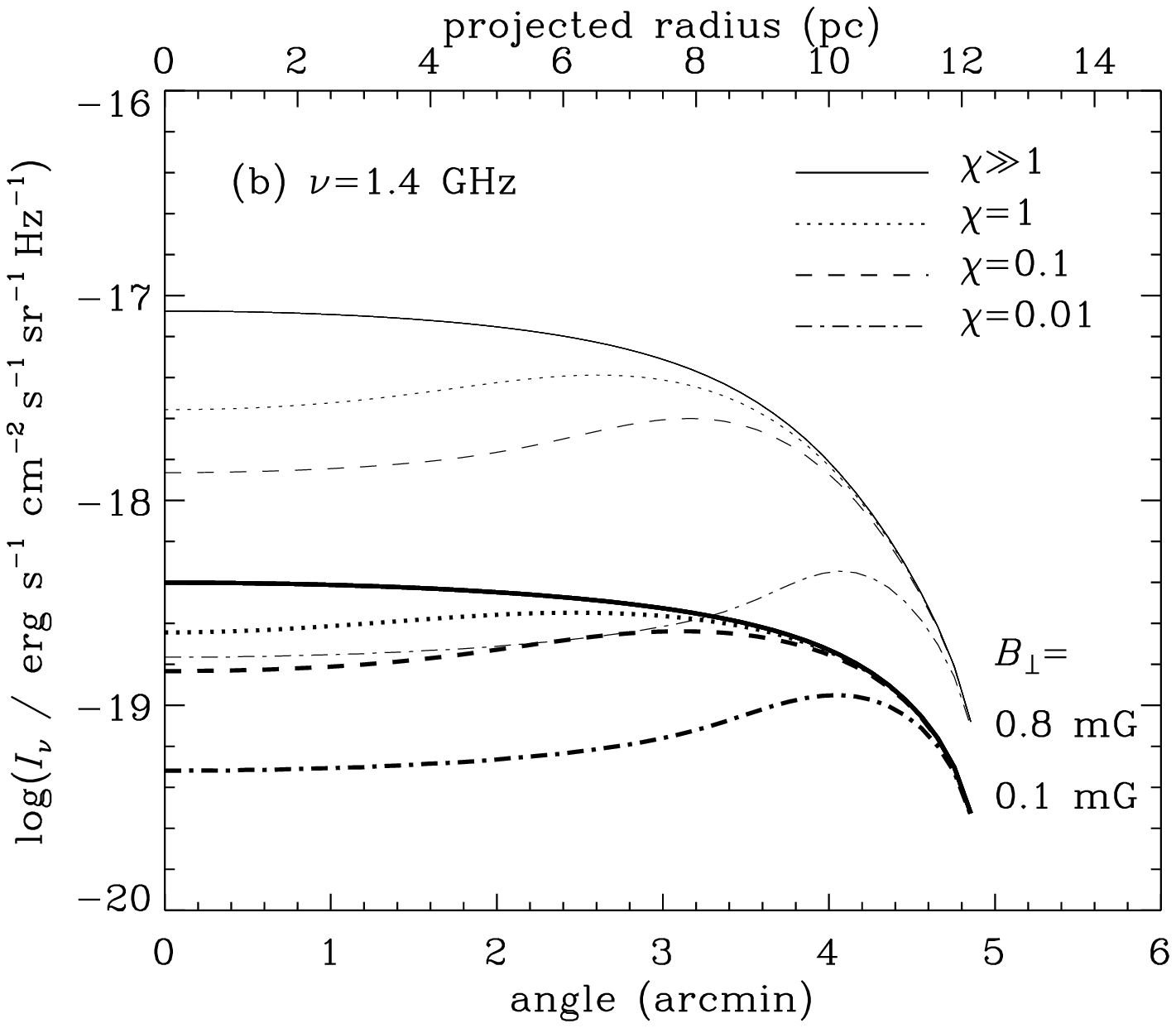}
\caption{ Predicted specific intensity 
$I_\nu(\theta)$ at (a) 330~MHz, and (b) 1.4~GHz, as a function of
angular distance from the centre of the Sgr~B2 complex for
diffusion as defined by Eq~\ref{Eq:diffco_Gabici} with $\chi$ as
labeled, a cosmic ray flux equal to that at Earth and for
magnetic field as labeled.}
\label{intensity_v_angle}
\end{figure}

We obtain the intensity $I_\nu(\theta)$, shown in
Fig.~\ref{intensity_v_angle} as a function of angular distance
$\theta$ for two frequencies, by integrating through the cloud
complex assuming the synchrotron emission is optically thin,
\[
I_\nu(\theta) =\int j_\nu(\vec{r}) d\ell.
\]
Depending
on the diffusive transport suppression factor $\chi$, we may
expect significant ``limb brightening'' of the synchrotron
emission.  We find that in the case of the Sgr~B2 complex most of
the flux comes from within $\sim$11~pc of its centre.

\section{Discussion}

Radio continuum observations of the Sgr~B2 GMC, including new
measurements at 1.4 GHz and 2.4 GHz, are discussed in detail in a
separate paper by \citet{JonesSgrB2MainCmplx2008}.  Since we want
to compare the continuum emission in the Sgr~B2 region with our
predictions for the region of the dense central region of the
giant molecular cloud we use continuum flux estimates for the
region 11~pc in size centred midway between Sgr~B2~(M) and
Sgr~B2~(N).  Note that other papers may use quite different sizes
for the Sgr~B2 region and hence quote very different fluxes.  In
this complex region of the Galaxy, when assembling radio spectra
from fluxes established by different groups, it is essential to
use fluxes obtained over the same solid angle.

\begin{figure}
\includegraphics[width=8cm]{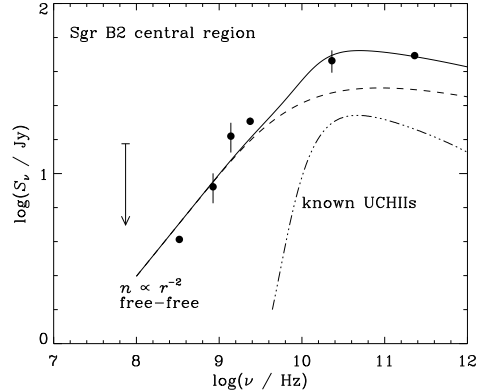}
\caption{Observed fluxes summarized by
  \citet{JonesSgrB2MainCmplx2008} from the central region of Sgr
  B2 complex including the major H{\sc ii} regions but excluding the
  Southern Non-Thermal Source.  The flux from the known UCHII
  regions is indicated (chain curve), and the best fitting model
  of free-free emission from a constant temperature spherical
  envelope or wind with $n \propto r^{-2}$ is shown by the
  dashed curve, and the solid curve gives the sum of the two
  thermal components.  }
\label{flux_thermal}
\end{figure}

No evidence was found for diffuse, non-thermal emission out to
$\sim$11~pc from the centre of the Sgr~B2 GMC with
limb-brightening as predicted for synchrotron emission by
secondary electrons in the previous section, consistent with the results of \citet{LangPalmerGoss2008} who find a thermal spectrum for Sgr B2 including its envelope.  Nor was there any
suggestion of polarised emission characteristic of synchrotron
emission.  However, the major H{\sc ii} regions Sgr B2(M) and Sgr
B2(N) showed up at all frequencies, and there is evidence of a
strong non-thermal source to the south, which we have marked as
``Non-Thermal Source'' in Fig.~\ref{sketch}, which is the subject
of a separate paper \citet{Jones2008b}.  Figure
\ref{flux_thermal} shows the spectral energy distribution for the
central region of the Sgr~B2 GMC (excluding the Non-Thermal
Source), and therefore includes the combined fluxes from the
major H{\sc ii} regions.

Sgr B2 is by far the most massive molecular cloud in the Galaxy
and it is near the Galactic Centre which is almost certainly a
region of enhanced cosmic rays.  When selecting Sgr~B2 for our
study, we knew that separating out the thermal emission would be
extremely difficult.  Our search for any comparable mass
molecular cloud with no star formation was unsuccessful.  

Before estimating upper limits to any synchrotron emission from
secondary electrons in the Sgr~B2 GMC, we shall attempt to fit the observed
spectral energy distribution solely by thermal emission processes.  A large
number of UCHII regions have been observed at high frequencies,
and so our first step will be to extrapolate
their spectra to low frequencies.  This will be done under the
assumption that each known UCHII region is a homogeneous sphere
of ionized interstellar gas, and we shall sum the contributions
from all known UCHII regions.  Clearly this will be an
approximation as there will also be
contributions from as-yet undiscovered UCHII regions, and because
many of the  UCHII regions will not be homogeneous, having density
gradients, e.g.\ winds.  We shall find that the UCHII regions
account for $\sim$50\% of the high frequency flux, and give a
negligible fraction of the observed flux at low frequencies under
these assumptions.  The second step will be to fit the residual
flux as thermal emission.  The shape of the spectral energy
distribution between 330~MHz and 1.4~GHz is suggestive of thermal
emission from a region, or regions, with a density gradient, and
we shall model it as a free-free emission from one, or many
identical, single temperature winds.  While this is clearly
unrealistic, the data available to us do not justify a more
sophisticated treatment.

\subsection{Free-free emission from UCHII regions}

The emission at the higher frequencies (22 GHz and 43 GHz) is
clearly thermal.  Regions with very low emission measures and
high 22~GHz and 43 GHz fluxes could potentially affect the
emission at low frequencies.  In order to investigate how much of
the flux at these lower frequencies could be attributed to UCHII
regions, we modeled the emission from the $\sim$60 known
individual compact and UCHII regions reported in
\citet{Gaume1995} and \citet{dePree1998}.  This was achieved by
``bootstrapping'' the flux at the respective frequencies such
that
\[
S_\nu = \sum_k
S_{\nu_i}^{(k)}\left({\nu\over\nu_i^{(k)}}\right)^2 \left({1-
e^{-\tau^{(k)}_\nu}\over 1- e^{-\tau^{(k)}_{\nu_i}} }\right)
\]
where $\nu_i=22$~GHz or 43~GHz, the sum over the $\sim$60 UCHII
regions with the label $(k)$ relates to the $k$th UCHII region,
and the frequency dependence of the thermal bremsstrahlung
absorption coefficient is taken from \citet{RybickiLightman1979}.
The summed flux from the known UCHII regions has been added to
Fig.~\ref{flux_thermal}, and it can be seen that the total flux from
these regions accounts for about 50\% of the 23 and 230 GHz flux
but their contribution below 3~GHz is negligible.  

\subsection{Free-free emission from envelopes or winds with density gradients}

Between 330~MHz and 1.4 GHz the spectrum may be fitted with a
single power-law $S_\nu \sim \nu^{0.6}$ characteristic of
optically thick emission from a spherical envelope or wind with
a density gradient of the form
\[
n_e(r)=n_i(r)= n_0 \left({r \over r_0}\right)^{-2}
\]
 as described by \citet{PanagiaFelli1975} who give the expected flux at low frequencies
\begin{eqnarray*}
S_\nu^{\rm thick}&=&0.611\left({n_0\over \mbox{1 cm$^{-3}$}}\right)^{4/3} \left({r_0\over\mbox{1~pc}} \right)^{8/3}\left({\nu\over \mbox{10~GHz}}\right)^{0.6}\\
 &&\left({T \over 10^4\mbox{~K}}\right)^{0.1} \left({d\over\mbox{1 kpc}}\right)^{-2}.
\end{eqnarray*}
For the case of  optically thin emission we can use
\[
\int_{r_0}^\infty 4\pi r^2 \,  n_e(r)\, n_i(r) \,dr = 4\pi 
{n_0}^2  r_0^3
\]
together with the free-free emission coefficient and Gaunt factor
$g(T,\nu)$ from \citet{RybickiLightman1979} to obtain the flux at
high frequencies where it is expected to be optically thin
\begin{eqnarray*}
S_\nu^{\rm thin}&=&1.6\times 10^{-5}\left({n_0\over\mbox{1 cm$^{-3}$}}\right)^{2}\left({r_0\over\mbox{1~pc}} \right)^{3}\\
&&\left({T \over 10^4\mbox{~K}}\right)^{-0.5} g(T,\nu)\left({d \over \mbox{1 kpc}}\right)^{-2}.
\end{eqnarray*}
Taking the optical depth to be $\tau_\nu= S_\nu^{\rm
thin}/S_\nu^{\rm thick}$ the flux from the envelope or wind is
$S_\nu=S_\nu^{\rm thick}(1 - e^{-\tau_\nu})$.  

We fit this density-gradient model to the observed 330~MHz to
230~GHz fluxes (after subtracting the contributions of known
UCHII regions as shown by chain curve in Fig.~\ref{flux_thermal})
and this is shown by the dashed curve in Fig.~\ref{flux_thermal};
the solid curve shows the sum of the two thermal components.  For
a temperature $T=10^4$~K the best fitting parameters are
$n_0=3.47\times 10^7$ cm$^{-3}$ and $r_0=4.12\times 10^{-3}$~pc.
The high density and small size would indicate that the emission
is likely to have come from winds off, or excited by, young stars
within the H{\sc ii} regions.  If the flux is due to $N$ separate
identical objects, their wind parameters would be $n_0=3.47\times
10^7 \times N$ cm$^{-3}$ and $r_0=4.12\times 10^{-3} / N$~pc in
order to give the same total spectrum.

\subsection{Synchrotron emission by secondary electrons }

\begin{figure}
\includegraphics[width=8cm]{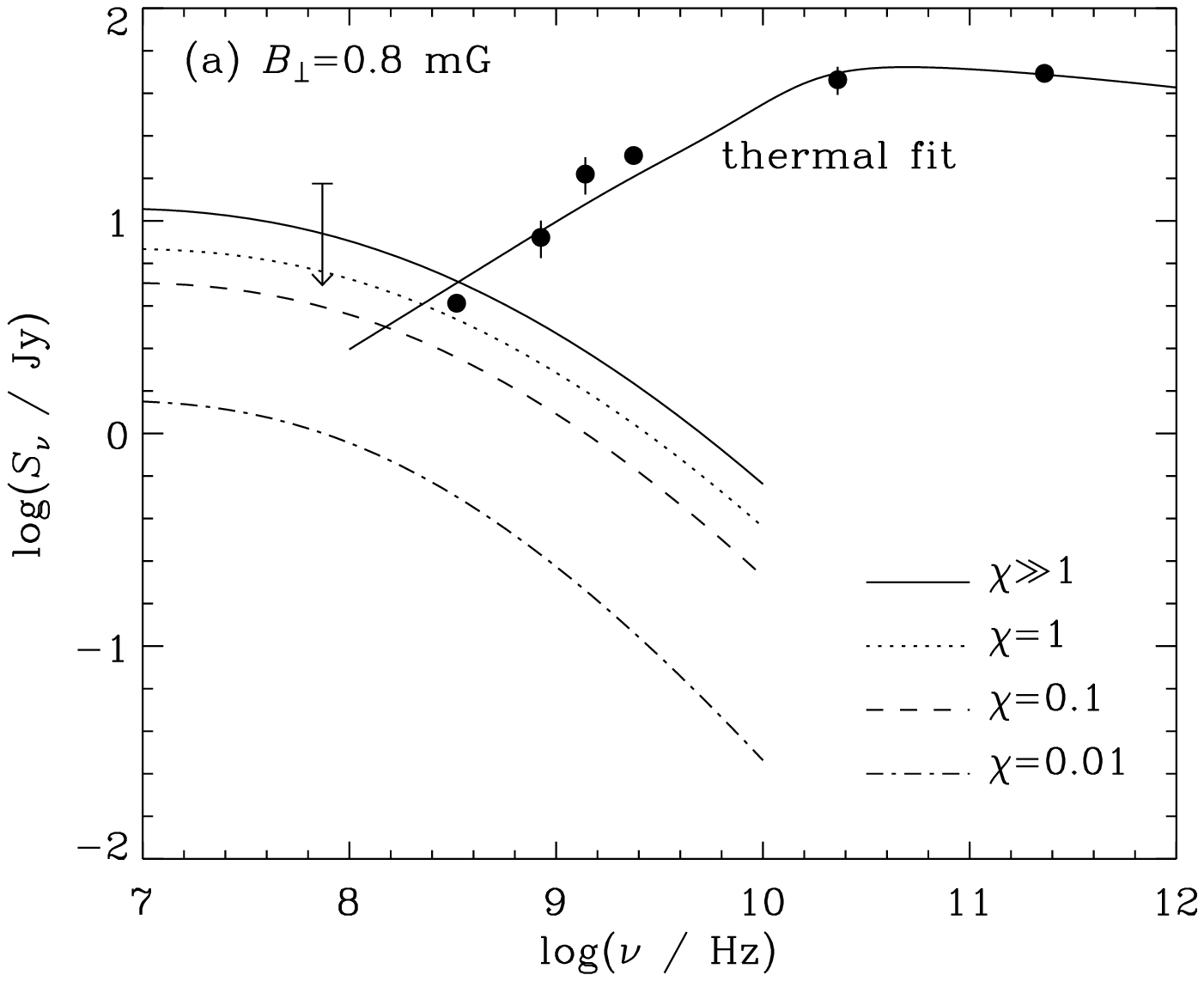}

\includegraphics[width=8cm]{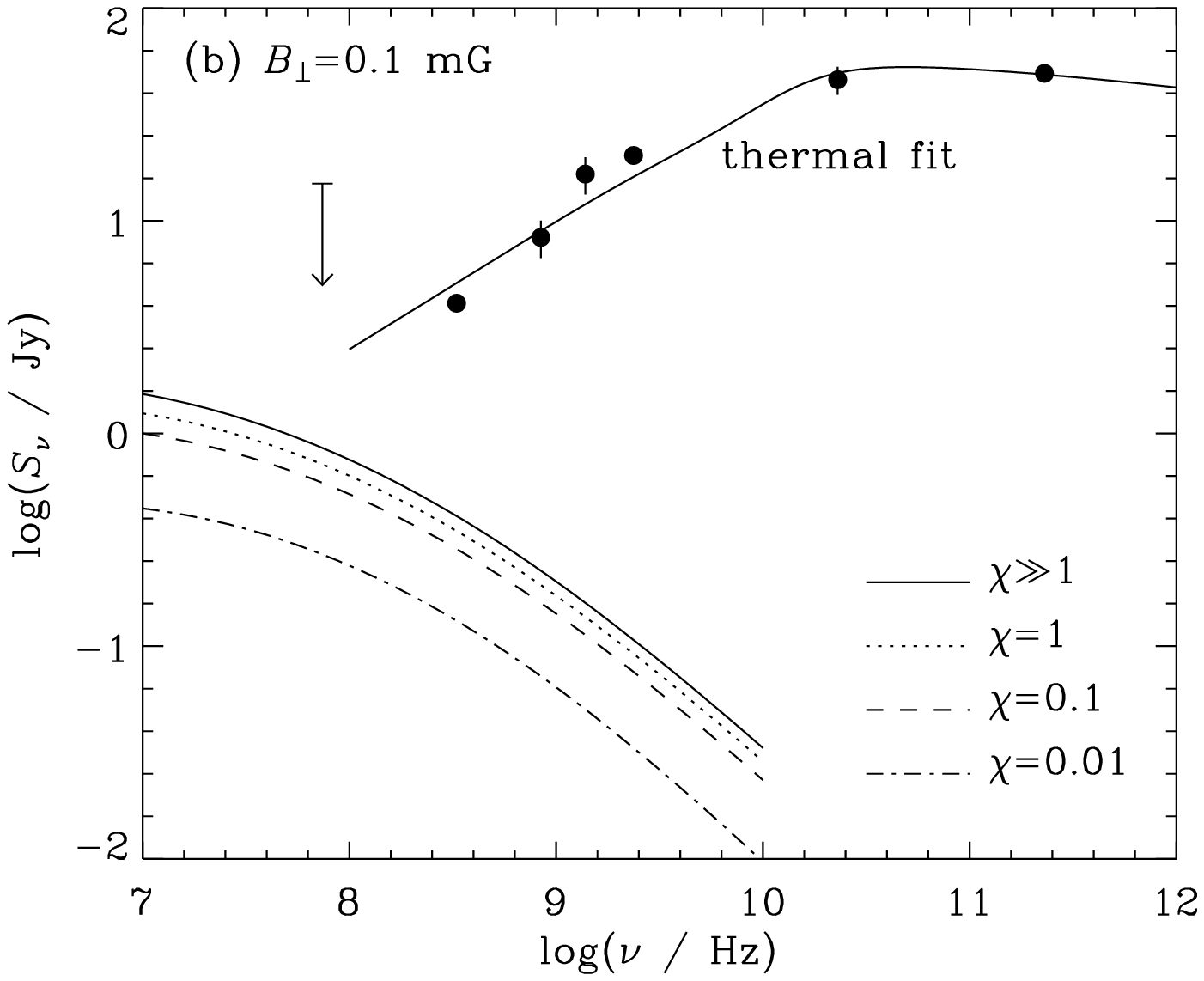}
\caption{Observed fluxes summarized by
\citet{JonesSgrB2MainCmplx2008} from the central region of Sgr B2
complex including the major H{\sc ii} regions but excluding the
Southern Non-Thermal Source plotted together with fluxes
predicted for (a) $B_\perp=0.8$mG, (b) $B_\perp=0.1$mG, and a
cosmic ray spectrum as in the solar neighbourhood.  Predicted
fluxes are shown for diffusion as defined by
Eq~\ref{Eq:diffco_Gabici} with $\chi$ as indicated.  The total
estimated thermal flux from the major H{\sc ii} regions is shown by the
upper solid curve.  }
\label{flux}
\end{figure}

In Fig.~\ref{flux} we re-plot the spectral energy distribution
and show the flux predictions for the synchrotron emission
from secondary electrons.  Given that we have found no evidence
of synchrotron emission from secondary electrons, and that the
observed radio continuum emission is consistent with a thermal
origin, we shall require that any flux of synchrotron emission
from secondary electrons be well below the observed 330~MHz flux.
For this, we shall somewhat arbitrarily adopt an upper limit of
$S_\nu$=1 Jy at 330 MHz for any non-thermal component.

Comparing the predicted synchrotron flux in Fig.~\ref{flux}(a)
with our adopted limit, we find that for $B_\perp$=0.8~mG
and $f_{CR}$=1, the diffusion factor must be 
$\chi<$0.02.  With a significantly lower magnetic field such as
$B_\perp$=0.1~mG (see Fig.~\ref{flux}b), even $\chi\gg 1$
(unimpeded cosmic ray penetration) is allowed and for this case
cosmic ray enhancement in the Galactic Centre region at multi-GeV
energies up to a factor $f_{CR}<$2.5 higher than in the Solar
neighbourhood is not excluded.  For this magnetic field, and more
reasonable diffusion factors, the maximum allowed cosmic ray
enhancement is 3 ($\chi$=1), 3.9 ($\chi$=0.1), 7.7 ($\chi$=0.01).
We emphasize that these are upper limits for cosmic ray
enhancement and that there is no evidence for any cosmic ray
enhancement at multi-GeV energies.  In fact for the higher
magnetic field, i.e.\ $B_\perp$=0.8~mG, which is based on
Zeeman splitting observations, the data suggest that there is no 
enhancement, or that cosmic rays at these low energies 
are unable to significantly penetrate into the Sgr~B2 GMC.

In conclusion, we have no evidence that synchrotron emission by
electrons and positrons produced by cosmic ray interactions has
been observed from the Sgr~B2 molecular cloud complex.  The most
likely explanation for this is that, for reasonable diffusion
models, cosmic rays with multi-GeV energies (that produce
secondary electrons with the right energy to radiate at GHz
frequencies in $\sim$0.8~mG fields) cannot penetrate into the
dense central regions of Sgr B2 GMC where much of the potential
mass of target nuclei is located.  This exclusion is also the
likely explanation for non-observation of the Sgr B2 GMC by EGRET
because it is again the same multi-GeV energy protons producing
pions in $pp$ collisions followed by $\pi^0\to \gamma\gamma$ that
make an important contribution to 100~MeV to multi-GeV
gamma-rays.  The observation of the Sgr B2 GMC by HESS
\citep{Aharonian2006} at TeV energies is consistent with more
complete penetration of cosmic rays at higher energies into the
dense central regions.  

In choosing giant molecular clouds in the central region of the
Galaxy for future investigation of their synchrotron emission by
secondary electrons, one would look for a GMC with with a mass of a few
$10^5$~M$_\odot$, a lower central density than Sgr~B2, e.g.\
$n_{H_2}$$\sim$$10^4$~cm$^{-3}$ so that low energy cosmic rays may
more easily penetrate it, a magnetic field above 0.1~mG and
little star formation. We do not know of any, but such clouds may
become apparent with the aid of new infrared surveys.

Finally, we emphasise that as we have no
independent knowledge of the diffusion coefficient within the
Sgr~B2 GMC, i.e.\ $\chi$ is unknown, we are unable to make a
definitive statement about the enhancement of the low energy
cosmic ray flux in the central region of the Galaxy relative to
that in the Solar neighbourhood.

\section*{Acknowledgments}
The National Radio Astronomy Observatory is a facility of the
National Science Foundation operated under cooperative agreement
by Associated Universities, Inc.  The Australia Telescope Compact
Array is part of the Australia Telescope which is funded by the
Commonwealth of Australia for operation as a National Facility
managed by CSIRO.  This research was supported under the
Australian Research Council's Discovery Project funding scheme
(project number DP0559991).  While this research was conducted
Professor R. D. Ekers was the recipient of an Australian Research
Council Federation Fellowship (project number FF0345330).  The
authors thank T.~Stanev for providing particle physics data
employed in this paper.  We thank the referee for helpful
comments.

\label{lastpage}

\end{document}